\documentclass[12pt,preprint]{aastex}  

\usepackage{wrapfig}
\usepackage{graphicx}
\usepackage{lscape}
\usepackage{epsfig}
\usepackage{epstopdf}

\newcommand{\eg}{{\it e.g.,~}}

\def\yesfig{1}

\def\figver{0}
\newcommand{\dofig}[1]{
\ifx\figver\yesfig
{#1}
\myclear 
\fi
}

\usepackage{color}
\newcommand{\myclear}{}

\begin{document}

\title {Galaxy Clusters at the Edge: Temperature, Entropy, and Gas Dynamics at the Virial Radius}

\author{ Jack~O.~Burns\altaffilmark{1}, Samuel~W.~Skillman\altaffilmark{1,2}, Brian~W.~O'Shea\altaffilmark{3}}

\altaffiltext{1}{Center for Astrophysics and Space Astronomy, Department of Astrophysical \& Planetary Science, University of Colorado, Boulder, CO 80309}
\altaffiltext{2}{DOE Computational Science Graduate Fellow}
\altaffiltext{3}{Department of Physics \& Astronomy and Lyman Briggs College, Michigan State University, East Lansing, MI, 48824}

\email{jack.burns@cu.edu}

\begin{abstract}

Recently, \textit{Suzaku} has produced temperature and entropy profiles, along with profiles of gas density, gas fraction, and mass, for multiple galaxy clusters out to $\sim$$r_{200}$ ($\approx$ virial radius).  In this paper, we compare these novel X-ray observations with results from N-body + hydrodynamic adaptive mesh refinement cosmological simulations using ~the ~\textit{Enzo} code.  There is excellent agreement in the temperature, density, and entropy profiles between a sample of 27 mostly substructure-free massive clusters in the simulated volume and the observed clusters.  This supports our previous contention that clusters have "universal" outer temperature profiles. Furthermore, it appears that the simplest adiabatic gas physics used in these \textit{Enzo} simulations is adequate to model the outer regions of these clusters without other mechanisms (e.g., non-gravitational heating, cooling, magnetic fields, or cosmic rays).  However, the outskirts of these clusters are not in hydrostatic equilibrium.  There is significant bulk flow and turbulence  in the outer intracluster medium created by accretion from filaments.  Thus, the gas is not fully supported by thermal pressure.   The implications for mass estimation from X-ray data are discussed. 

\end{abstract}


\section{Introduction} \label{sec:intro}
Galaxy clusters are unique and potentially powerful cosmological probes of the Universe.  They are the largest gravitationally bound objects, having grown hierarchically within the large-scale cosmic web.  As such, clusters are an important source of information about the components of the universe in which they formed and evolved \cite[\eg][]{voitrev05, 2009arXiv0906.4370B}.  Gravity drives structure formation within an expanding universe, with regions of density higher than the average becoming gravitationally bound and decoupling from the expansion.  Clusters probe the high density tail of this cosmic density field.  The number density of clusters is strongly dependent upon the specific cosmological model, especially when viewed as a function of redshift \cite[\eg][]{rosati}.  The mass function of clusters and its evolution are also sensitive to $\sigma_8$ (a parameter that quantifies the RMS density fluctuations on comoving scales of 8 h$^{-1}$ Mpc).  The potential to use the cluster mass function to measure cosmological parameters is challenging observationally because mass is not a direct observational quantity. Rather, X-ray luminosity, X-ray temperature, thermal SZE (Sunyaev-Zeldovich Effect) integrated $Y (\propto \int \rho T dl)$, or weak lensing shear are the observables which must then be translated into an estimate of cluster mass.

Over the past decade, X-ray observations from {\it ROSAT}, {\it Chandra}, {\it XMM-Newton}, and, more recently {\it Suzaku}, have begun to produce significant samples of clusters out to $z \approx 1.3$ \citep[\eg][]{2008MNRAS.383..879A}.  With these samples, it is becoming possible to distinguish between cosmological models using the evolution of the cluster mass function \cite[\eg][]{2009ApJ...692.1060V} and the cluster gas fraction \cite[\eg][]{2008MNRAS.388.1265R}.  The primary limiting factor in such applications remains accurate conversion between X-ray observables such as X-ray luminosity and temperature to cluster mass.  This is complicated by the nonlinear baryonic processes at the cores of clusters, including radiative cooling and non-gravitational heating from supernovae and active galactic nuclei \cite[AGNs; \eg][]{rusz02, heinz98, burns98} and possibly nonthermal processes involving cosmic rays \cite[\eg ~see][]{2008ApJ...689.1063S} and magnetic fields \cite[\eg ~see][]{2008MNRAS.388.1062C, 2009ApJ...698L..14X}. Other complications include possible bias and scatter in scaling relations \citep{motl05, nagai}, as well as errors created by assuming that cluster gas is in hydrostatic equilibrium \cite[\eg ~see][]{2009ApJ...705.1129L, burns08, rasia06, nagai07}.  

Because of the nonlinear nature of structure formation, the regions outside the cores do not necessarily behave in this same complex manner \citep[\eg][]{2006MNRAS.373.1339R}.  The peripheries of clusters have lower gas and galaxy densities, and long cooling times comparable to the Hubble time, thus potentially making the thermodynamics simpler than in the cores.  At the same time, the outer regions of clusters are closer to the sources of accretion from filaments and, therefore, the gas dynamics may be more complex.

The intracluster X-ray gas in galaxy clusters is often assumed to be in hydrostatic equilibrium which relates the gravitational potential ($\Phi$) to the gas pressure ($P$) and the gas density ($\rho$) such that
\begin{equation}
\nabla \Phi = -\frac{\nabla P}{\rho}.
\end{equation}
Applying Gauss's Law to the gravitational potential in the above, the cumulative mass is
\begin{equation}
M(<r) = \frac{1}{4 \pi G} \int -\frac{\nabla P}{\rho} dA,
\end{equation}
where G is Newton's gravitational constant and the integral is over a spherical surface area with radius $r$.  If the cluster is further assumed to be spherically symmetric and the pressure arises only from thermal motions ($P = {\rho kT}/[\mu m_p])$, then
\begin{equation}
M(<r) = -\frac{r^2 k}{\rho G \mu m_p} \left[T \frac{d \rho}{dr} + \rho \frac{dT}{dr} \right].
\end{equation}
As seen in Equation (3), we need to measure both the gas densities and temperatures, along with their gradients, to calculate the hydrostatic equilibrium masses.  Gas density profiles can be accurately determined from the X-ray surface brightness, $S_X$, since $S_X \propto \rho^2$ with only a weak dependence on temperature.

Temperature profiles, on the other hand, are more challenging since they require spatially-resolved X-ray spectroscopic observations acquired from relatively long integrations.  This has been difficult, particularly to measure cluster temperatures beyond $\sim 0.5 r_{200}$\footnote{$r_{200}$ is the radius at which the density is 200 times critical and $\approx$ virial radius for a flat $\Lambda$CDM Universe \citep{1998ApJ...495...80B}}.  Because of the low and stable particle background levels at its orbit, {\it Suzaku} has begun to change the landscape by producing X-ray temperature profiles out to $\approx$$ r_{200}$ for a small number of clusters including PKS 0745-171 \citep{George2009}, Abell 1795 \citep{Bautz2009}, Abell 399/401 \citep {Fujita2008}, Abell 2204 \citep{Reiprich2009}, and Abell 1413 \citep{2010arXiv1001.5133H}.  In each cluster, the temperature is observed to decline by a factor of $\approx 3$ from the peak near the cluster core to regions at $\approx r_{200}$.

Using the Eulerian adaptive mesh refinement cosmology code {\it Enzo} for a $\Lambda$CDM universe, our group first proposed a universal temperature profile for galaxy clusters \citep{Loken2002}.  This average profile was well-fit by a power-law out to the virial radius.  This profile agreed well with the X-ray data available at that time for nearby clusters from {\it BeppoSAX} \citep{Degrandi2002}. Subsequent {\it Chandra} observations also appear consistent with such a universal temperature profile beyond the dense central cores for galaxy clusters \citep[\eg][]{VikhlininKravtsov2006}. However, these observed X-ray temperature profiles extended out only to $\approx 0.5 r_{200}$ so it was unclear if the universal temperature profiles for numerical clusters are in agreement with the outer profiles for real clusters.

In this paper, we compare the temperature, density, and entropy profiles for a new sample of numerical clusters generated using the {\it Enzo} cosmology code with new observations from {\it Suzaku}.  We ask the question: Do real clusters follow a universal temperature profile out to the virial radius, as is predicted by numerical simulations with simple adiabatic gas physics?  We also explore the implications of this particular form of the universal temperature profile for hydrostatic equilibrium, intracluster gas dynamics generated by on-going accretion in the outer periphery of clusters, and cluster mass estimation.

In Section \ref{sec:simulations}, we describe the {\it Enzo} cosmology code and the numerical simulations used for the comparison with X-ray observations.   In Section \ref{sec:results}, we compare gas densities, temperatures, and entropy profiles between our numerical simulations and observed clusters.  Then, in Section \ref{sec:discussion}, we explore the implications of the form of the observed and simulated cluster temperature profiles on hydrostatic equilibrium and cluster mass determinations.  We end with a summary and conclusions in Section \ref{sec:summary}.

\section{Simulations}\label{sec:simulations}

We use the \textit{Enzo} adaptive mesh refinement (AMR) N-body + hydrodynamics
cosmology code \citep{Bryan:1997aa,Bryan:1997ab, Norman:1999aa,
  OShea:2004aa, 2005ApJS..160....1O} to simulate a comoving volume of $(128\
h^{-1}\mathrm{Mpc})^3$ with $256^3$ root-grid cells and up to 5 levels
of additional refinement.  This simulation utilizes the ZEUS
finite-difference method \citep{Stone:1992aa,Stone:1992ab}. The AMR is
controlled by refining any region that is overdense by a factor of 8
in either the dark matter or gas density.  The peak resolution is
$15.6~h^{-1}\mathrm{kpc~(comoving)}$.  The initial conditions are
generated from an \citet{Eisenstein:1999aa} power spectrum with a primordial
spectral index $n_s=0.97$.  We use the cosmological parameters
$(h,\Omega_M,\Omega_{B},\Omega_{CDM}, \Omega_{\Lambda}, \sigma_8) = (
0.704, 0.268, 0.0441, 0.2239, 0.732, 0.9)$, with $h= H_0/(100\
\mathrm{km}\ \mathrm{s}^{-1}\mathrm{Mpc}^{-1})$.  The simulation was
initialized at $z=99$ and run until $z=0$.  The dark matter mass
resolution is $3.12\times10^9~h^{-1}M_\odot$.  For a further exposition of Enzo and its use in studying the statistical properties of clusters, see \citet{2008ApJ...689.1063S}.

For all the analyses of the simulations and observations throughout this paper, we will assume $H_0 = 70$ km/sec/Mpc and the other cosmological parameters given above.

\section{Comparison of Simulated and Observed Clusters}\label{sec:results}

\begin{figure}
\begin{center}
\vspace{-3.5cm}
\includegraphics[width=6.5in]{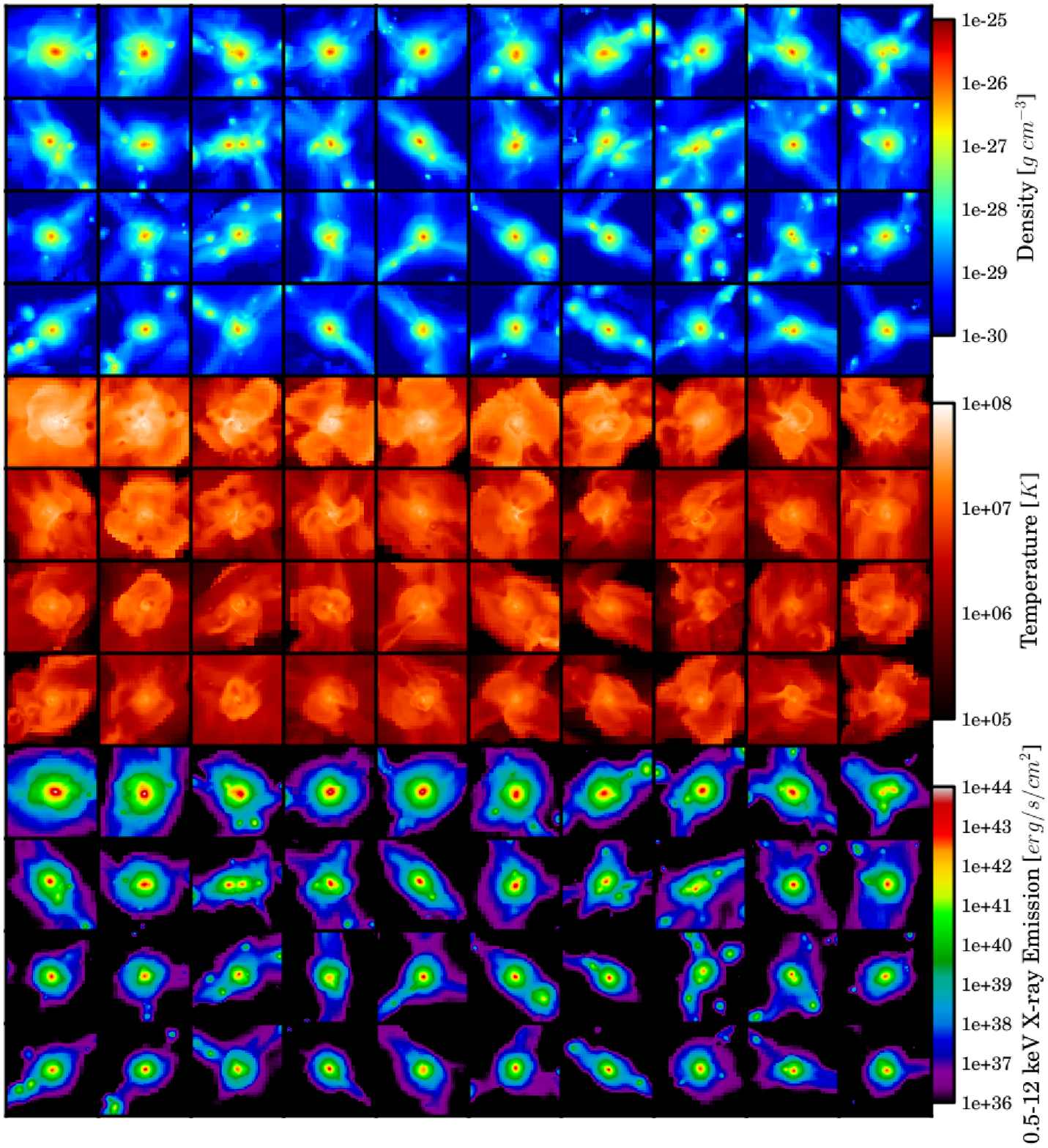}
\caption{Grid of images of the 40 most massive clusters in the numerical sample selected from a cosmological volume at $z=0.05$. The images are ordered by $M_{200}$ from the upper left ($10^{15}$ M$_{\odot}$) to the lower right ($1.7 \times 10^{14}$ M$_{\odot}$) for each of the three panels of gas density (top), temperature (middle), and X-ray surface brightness (bottom).  Field of view of each image is 8 $h^{-1}$ Mpc.  The range of properties for these clusters is listed in Table 1. This sample was reduced to 27 clusters for statistical analysis by eliminating obvious merger systems.  }

\label{ImageGrid}

\end{center}  
\end{figure}

We selected the 40 most massive clusters in our numerical volume at $z=0.05$ for
comparison to the \textit{Suzaku} observations because these clusters best match
those observed (Table \ref{tab:properties}).  This sample was then edited by examining the 3D distribution of the dark matter.  Any cluster with a 
secondary dark matter density peak located outside of $0.1 r_{200}$ was marked as ``disturbed''.  Upon visual inspection, these clusters were
confirmed to be ones with major substructure.  This produced a
``clean'' sample of the 27 most massive clusters that are relatively
substructure-free.  However, a note of caution is warranted, as there
is accretion of smaller halos (subclusters) present in nearly all
clusters; such small halo accretion is better seen in temperature images where shocks from supersonically merging subclusters are more obvious (see Fig. \ref{ImageGrid}).

\begin{table}[t]
\begin{center}
\caption{Comparison of Properties of Simulated with Observed Clusters}
\label{tab:properties}
\begin{tabular}{|p{4cm}|p{2cm}|p{3cm}|p{3cm}|}
\hline
\hline
{\bf \small{Cluster}} & $z$ & $M_{200}$ (M$_{\odot}$) & $r_{200}$ (Mpc) \\
\hline
PKS 0745-191 & 0.103 & $6.4 \times 10^{14}$ & 1.72 \\
\hline
Abell 1795 & 0.063 & $8.6 \times 10^{14}$ & 1.9 \\
\hline
Simulated Clusters & 0.05 & $1.7-10 \times 10^{14}$ & $1.6-2.8$ \\
\hline
\hline
\end{tabular}
\vspace{-3mm}
\end{center}
\end{table}

A grid of images of gas densities, temperatures, and $0.5-12.0$ keV
X-ray emission is portrayed in Figure \ref{ImageGrid}.  For each temperature and density projection, we show a corresponding X-ray image.  The method of \citet{Smith:2008aa} was used for the radiative cooling to calculate frequency-dependent emission using the emission function from the Cloudy code
\citep{Ferland:1998aa}.  Given a temperature and density of the gas, and assuming a metallicity of 0.3 solar, this returns the total X-ray emissivity.  The X-ray surface brightness is the integral along the line of sight of the X-ray emissivity.  These images in Figure \ref{ImageGrid} show a good deal of non-circularly symmetric structures, often produced by accretion along filaments, as also appears to be seen in the clusters observed by \textit{Suzaku} \citep[e.g.,][]{George2009, Bautz2009}.

\clearpage

\clearpage

\subsection{Analysis Pipeline}\label{pipeline}
\subsubsection{3D Profiles}
For each cluster, we create spherically averaged radial profiles centered on the center of mass of the cluster.  We track the
following key fields: radius (in Mpc), overdensity of a sphere out to the
current radius with respect to the background density ($\Omega_m \rho_{cr}$), dark matter density, gas density, entropy, pressure, specific radial kinetic energy, specific kinetic energy, temperature, thermal energy, X-ray emission, total energy, and enclosed baryonic + dark matter mass.  

An important point in our profiling procedure involves the various
options for weighting the profiles.  In this study, we use two
different methods.  First we can use the simple method of weighting by
the mass of the cell.  Therefore, regions in radial shells in
high-density regions will be weighted higher than those in underdense
regions.  Second, we weight by the total X-ray emission emanating from
each cell.  This biases towards overdense regions even more strongly
since the X-ray emissivity scales roughly with $\rho^2$.  In Figures
3 and 4, we use X-ray weighting to compare with X-ray observables.  In
Figure 5, we use mass weighting to investigate the physical causes of
deviations from hydrostatic equilbrium.

\subsubsection{Profiles of 2D Projections}

Because observations yield inherently 2D images, it is important to
compare projections of our clusters to the true observations.  We
do this by first creating halo projections of each cluster that
encompass an 8 Mpc cube around the individual cluster.  For each of
these projections, we again have a choice between weighting by the
density or X-ray emission.  Unless otherwise noted, we will use only
the projections weighted by the X-ray emission.  From each of the 2D
projections, we then seek to create radial profiles.  Here we give
details of creating 2D profiles of X-ray weighted temperature, and use
the same technique in all other quantities.  

\begin{figure}[t]
\centering
\vspace{-8.5cm}
\includegraphics[width=6.5in]{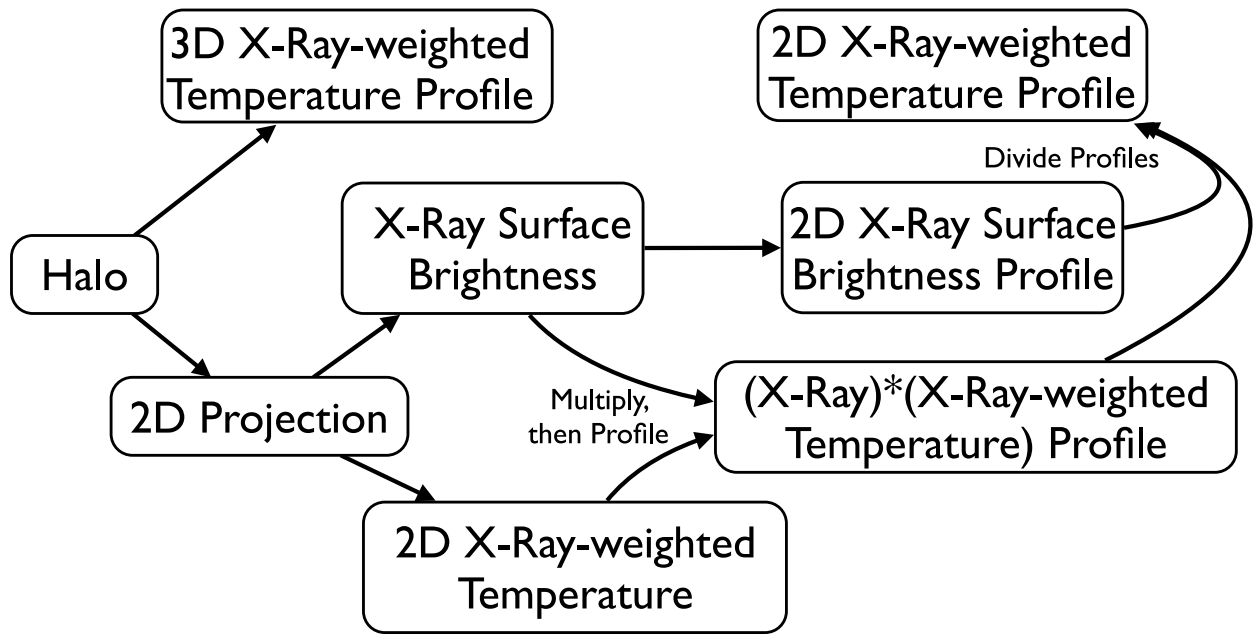}
\caption{Analysis Pipeline to construct both a 2D and 3D X-ray weighted profile of a halo.}
\label{fig:flowchart}
\end{figure}

For each cluster, we first make a projection of the X-ray emissivity, $\epsilon_X$,
which yields the X-ray surface brightness,
\begin{eqnarray}
  S_{X, Proj}(x,y) = \int \epsilon_X(x,y,z)dz.
\end{eqnarray}
  We then make a projection of the temperature, weighted by X-ray emissivity,
\begin{eqnarray}
 T_{Proj,X-Ray-weighted}(x,y) = \frac{\int T(x,y,z) \epsilon_X(x,y,z) dz}{S_{X, Proj}(x,y)}.
\end{eqnarray}
From these quantities, we create the profiles of the 2D projections by weighting the temperature projection by the surface brightness,
\begin{eqnarray}
 T_{2D, X-Ray-weighted}(r) = \frac{\int T_{Proj,X-ray weighted}(x,y) S_{X, Proj}(x,y) r d\theta }{\int S_{X, Proj}(x,y) r d\theta},
\end{eqnarray}
where $r$ and $\theta$ are the normal polar coordinates.  This results
in a correctly weighted radial profile of X-ray weighted temperature.
Note that this is different than simply creating a 2D profile from the
projection of X-ray weighted temperature. We can compare this final
step to the equivalent step in the creation of the spherically
averaged radial profile of the same quantity using normal spherical coordinates,
\begin{eqnarray}
  T_{3D, X-Ray-weighted}(r) = \frac{\int\int T(x,y,z) \epsilon_X(x,y,z) r^2 sin \theta d\theta d\phi }{\int\int \epsilon_X(x,y,z) r^2 sin \theta d\theta d\phi}.
\end{eqnarray}
For an illustration of this pipeline, see Figure \ref{fig:flowchart}. 

\clearpage
\clearpage

\subsection{Results}

\begin{figure}
\begin{center}
\vspace{-1.5cm}
\includegraphics[width=6.5in]{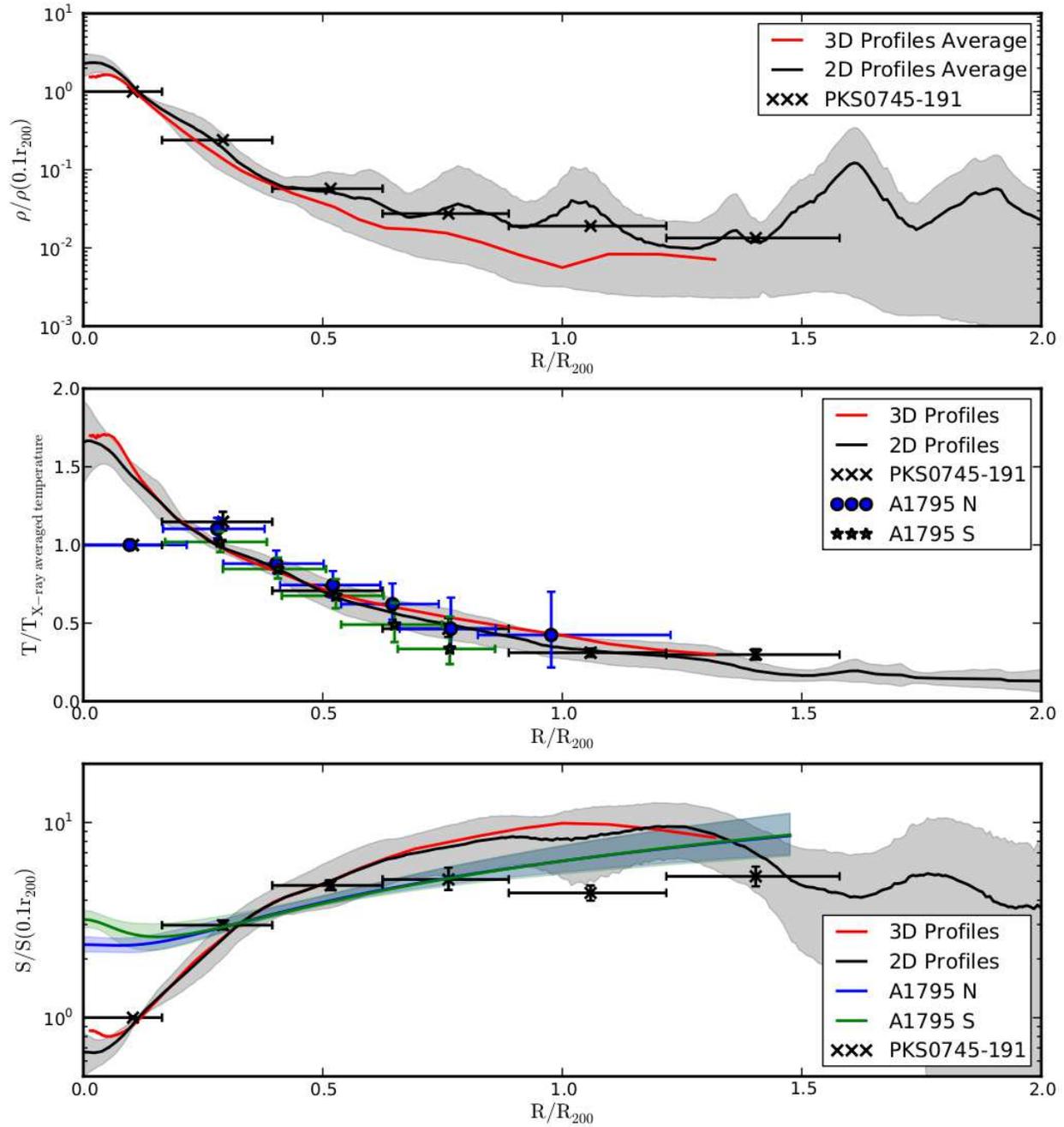}

\caption{3D and 2D X-ray weighted average profiles, as described in Section \ref{pipeline}, of normalized density, temperature, and entropy for simulated clusters.  The shaded regions are the standard deviations of each quantity produced from the scatter within the sample of numerical clusters.  The \textit{Suzaku} data points for PKS 0745-191 and A1795 (measured for two separate analysis regions, north (N) and south (S) of the core) are shown for comparison.  The blue and green shaded regions in the bottom entropy profile are for the N and S analysis regions of A1795. The temperature is normalized by the average value between 0.2 to 1.0 $r_{200}$, whereas the density and entropy are normalized by values at 0.1 $r_{200}$ and 0.3 $r_{200}$, respectively.  This difference in normalization produces a slight shift in the normalized entropy between simulations and PKS 0745-191.}

\label{3plots}

\end{center}  
\end{figure}

Applying the approach described in Section \ref{pipeline} to the analysis of the simulated clusters, we produced radial profiles of density, temperature, and entropy as shown in Fig. \ref{3plots}.  Here, we define entropy as $K \equiv kT (\frac{\rho}{\mu m_p})^{-2/3}$. The profiles were constructed by binning in circular annuli centered on the peak X-ray emission.  The temperature profiles were each normalized to their X-ray weighted average values between 0.2 to 1.0 $r_{200}$ so we could easily compare clusters of different masses. Such normalization is less effective for density and entropy which were, instead, normalized by values at 0.1 $r_{200}$ and 0.3 $r_{200}$, respectively. The shaded regions illustrate the standard deviations in quantities derived from the sample of numerical clusters.  This scatter generally grows with radial distance, especially for the density and entropy, beyond $\approx 0.7 r_{200}$.  This scatter reflects real variations from cluster to cluster in the outer regions produced by non-spherically symmetric accretion along filaments as mentioned above.  The flat outer profile ($>0.5 r_{200}$) seen for the average entropy of the numerical cluster sample in Figure \ref{3plots} is, in part, due to the averaging of individual profiles, some with positive and some with negative slopes.  

The average temperature profile shown in Figure \ref{3plots} is well-fit by a function of the form
\begin{equation}
\frac{T}{T_{avg}} = (1.74 \pm 0.03) {\left[ 1 + (0.64 \pm 0.10) \left(\frac{r}{r_{200}}\right) \right]}^{-3.2 \pm 0.4}.
\end{equation}
As we discuss in Section \ref{sec:discussion}, the particular shape/slope of this power-law for the temperature profile has implications for understanding the gas dynamics and ICM pressure support in the outer regions of clusters.

In Fig. \ref{3plots}, we also compare the numerical profiles with observed ones for two nearby clusters (PKS 0745-191 and A1795) with recent {\it Suzaku} X-ray measurements that extend out to $\sim r_{200}$.  A summary of the characteristics of these two observed clusters, along with the average properties of the numerical clusters, is given in Table \ref{tab:properties}.  Both observed clusters have small cool cores.  Since our simulations were purposely constructed with simple adiabatic gas physics, we did not expect to match the central regions of these clusters.  Instead, we focus on the outer profiles, beyond $\gtrsim 0.5 r_{200}$.  Other clusters recently observed by \textit{Suzaku} have similar outer temperature profiles with a factor of $\approx 3$ decline in temperature between the core and $r_{200}$ \citep[\eg][]{Reiprich2009, 2010arXiv1001.5133H}. 

\clearpage
\clearpage

It is clear from this figure that the outer cluster radial distributions of density, temperature, and entropy for PKS 0745-191 and A1795 agree well with the simulated clusters.  From Figure \ref{3plots}, we draw three conclusions about the regions outside the cluster cores.  First, the observed clusters follow ``universal'' temperature, density, and entropy profiles as characterized by numerical simulations out to the outer bounds of the clusters ($r_{200}$).  Second, the simplest intracluster medium (ICM) gas physics with an adiabatic equation of state is sufficient to characterize the thermodynamic gas properties of the peripheries of these clusters.  With the present \textit{Suzaku} X-ray data, added gas physics such as cooling, non-gravitational heating (via \eg low power radio or X-ray AGNs associated with non-central cluster galaxies \cite[\eg][]{2009ApJ...705..854H}) or nonthermal pressure (due to ICM B-fields and cosmic rays \cite[\eg][]{2008ApJ...689.1063S}) are not required, as also found by \citet{2006MNRAS.373.1339R}.  Third, the nonspherical and dynamical gas/halo accretion environment present in the outer reaches of the numerical clusters is important in reproducing the observed properties, as is discussed in the next section.

\section{Discussion}\label{sec:discussion}
\subsection{Are Galaxy Clusters in Hydrostatic Equilibrium?}\label{sec:equilibrium}

What does the particular shape of the observed and numerical cluster temperature profiles imply about the dynamical state of the ICM?  In particular, is this universal temperature profile consistent with the simplest form of hydrostatic equilibrium given in Equation (3), where the ICM pressure is strictly thermal?

\begin{figure}
\begin{center}
\vspace{-3.5cm}
\includegraphics[width=6.0in]{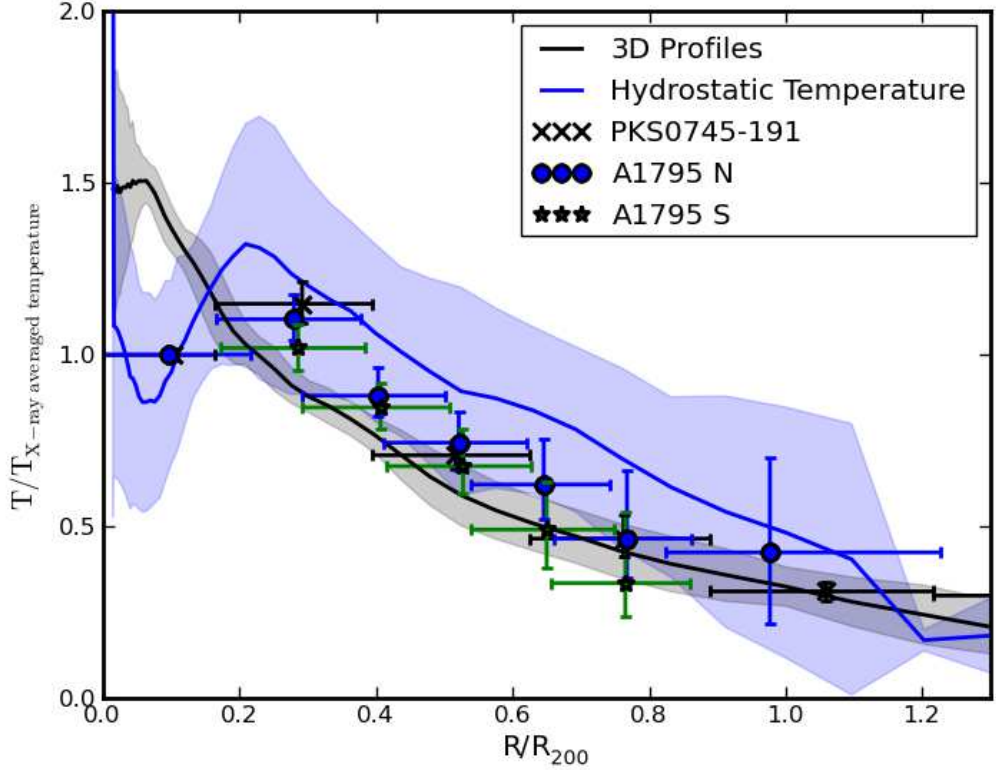}
\caption{Comparison of numerical and observed cluster temperature profiles with that expected for hydrostatic equilibrium.  The blue line and blue shading are the temperature distribution and standard deviations, respectively, expected if the gas in the dark matter potential wells of simulated halos is in hydrostatic equilibrium.  The black line and grey shading are the X-ray weighted average temperature and standard deviations, respectively, for the numerical clusters.  \textit{Suzaku} data points are also shown for PKS 0745-191 and for two analysis regions (north and south) in A1795.}

\label{comparetemp}
\end{center}
\end{figure}

In Figure \ref{comparetemp}, we compare the temperature profiles from the observed clusters and the simulations with that expected if the numerical clusters are in hydrostatic equilibrium.   To calculate the hydrostatic temperature, we begin by making spherically-averaged 3D radial profiles of the dark matter density, weighted by the X-ray emission.  From these radial profiles, we calculate the total enclosed mass as a function of radius.  We then assume hydrostatic equilibrium to calculate the pressure derivative at each point,
\begin{equation}
  dP = -\frac{G M(r) (\epsilon\rho_{dm})}{r^2} dr,
\end{equation}
where $\epsilon = \rho_{gas}/\rho_{dm}$ is the ratio of gas to dark matter density and is assumed to be constant. We then integrate the pressure inwards, assuming $P=0$ at our outermost radial point.  Finally, we calculate the temperature from 
\begin{equation}
  T_{hydro} = \frac{\mu m_p P }{k \epsilon \rho_{dm}}.
\end{equation}
Because we normalize the temperature by the X-ray weighted average temperature between $0.2 r_{200} < r < r_{200}$, the absolute scaling of the pressure (and the value of $\epsilon$) does not affect the result.  From the set of each individual profile, we then calculate the mean and standard deviation shown in Figure \ref{comparetemp}.  

Using a $\chi^2$-test to compare the temperatures and their standard deviations for the simulated clusters in Figure \ref{comparetemp}, we find that the numerical 3D radial temperature profile and $T_{hydro}(r)$ have a 1.6\% probability of being drawn from the same parent distribution.  Although the scatter is large, the average hydrostatic temperature profile is $\approx 20$\% higher than the 3D numerical clusters or observed \textit{Suzaku} clusters temperature profiles.  This raises concerns about accurately calculating cluster masses using this method, especially for applications to precise cosmological parameter estimation.

\begin{figure}
\begin{center}
\vspace{-3.5cm}
\includegraphics[width=6.0in]{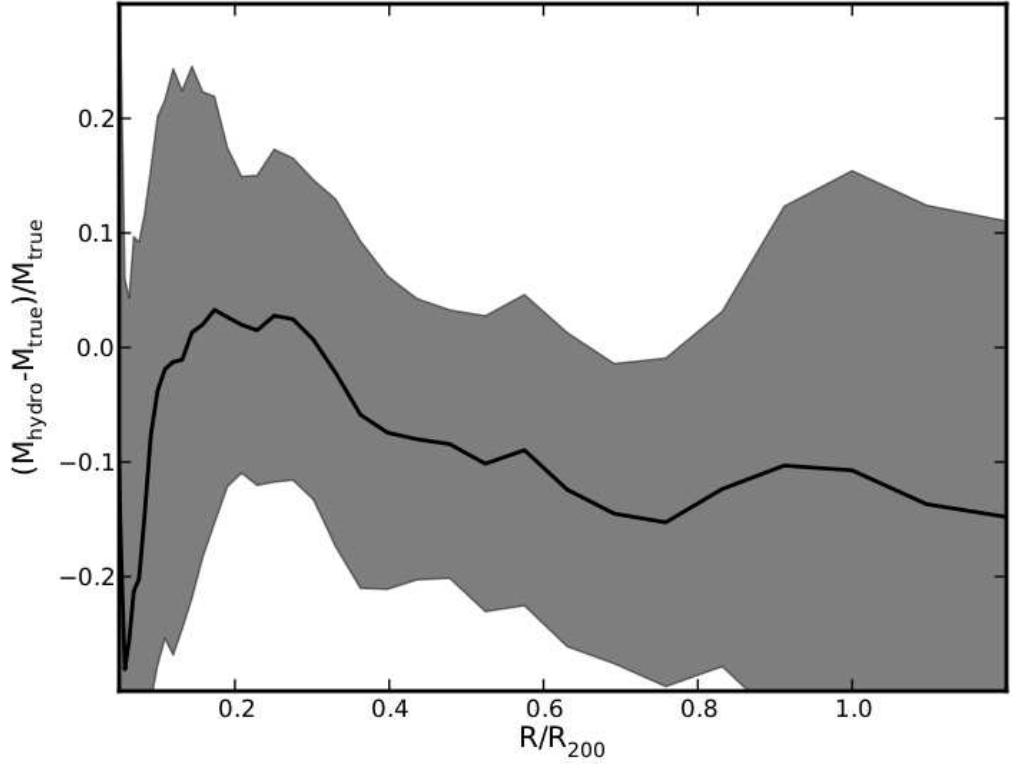}
\caption{Radial profile of the fractional difference between cluster mass assuming hydrostatic equilibrium and the actual mass for the sample of numerical clusters.  The shaded region is the standard deviation around the mean values.  Note the increasing underestimate of mass with radial distance beyond $\approx$0.3$r_{200}$.}

\label{hydromass}
\end{center}
\end{figure}

In Figure \ref{hydromass}, we investigate this issue further by plotting $(M_{hydro}(<r) - M_{true}(<r))/M_{true}(<r)$, where $M_{hydro}(<r)$ is the mass within a sphere with radius $r$ calculated from hydrostatic equilibrium using Equation (3) and $M_{true}(<r)$ is the dark matter mass.  In this case, we use a mass weighting to probe the physical cause of the differences in mass.  Previous numerical simulations have shown that there is a systematic bias (underestimate) of 5-15\% in the calculated mass of the central to mid regions of clusters ($r_{2500} - r_{500}$) \citep{burns08, rasia06, 2009ApJ...705.1129L}  assuming hydrostatic equilibrium.  Fig. \ref{hydromass} illustrates that this bias increases towards the edge of clusters. Beyond $\approx 0.8 r_{200}$, the average integrated mass is biased low by $\approx$$15$\%.  Recent observational work by \cite{2008MNRAS.384.1567M} supports this result in that they find hydrostatic cluster mass estimates are systematically low by 10-20\% compared to masses calculated via a weak lensing analysis out to $r_{500}$.  Importantly, Figure \ref{hydromass} also illustrates the large scatter in the hydrostatic equilibrium calculation due to different dynamical states of individual clusters (even after we edited the sample for obvious mergers).  At $0.8 r_{200}$, cluster hydrostatic masses range from 5\% overestimates to 30\% underestimates.

\clearpage
\clearpage

\clearpage

\subsection{Gas Dynamics in the Outer Regions of Clusters}\label{sec:dynamics}

Why does the intracluster medium of galaxy clusters not follow the simple prescription for hydrostatic equilibrium shown by Equation (3)?  There are several obvious reasons which relate to the connection of clusters with the cosmic web and to their on-going evolution via accretion of gas and subclusters from filaments.

First, unlike the view of several decades ago, galaxy clusters are not simple spheres of gas and dark matter that are disconnected from their surroundings.  Rather, they are closely tied to linear filaments within the large scale structure of the Universe.  Gas, galaxies, and dark matter are funneled along these filaments into clusters, which typically lie at the intersections of the filaments.  As a result, accretion onto clusters is complex and nonspherical.  Thus, azimuthal variations in $\nabla$$P$ using Equation (2) need to be folded into the calculation of $M(<r)$.

\begin{figure}
\begin{center}
\vspace{-1cm}
\includegraphics[width=6.0in]{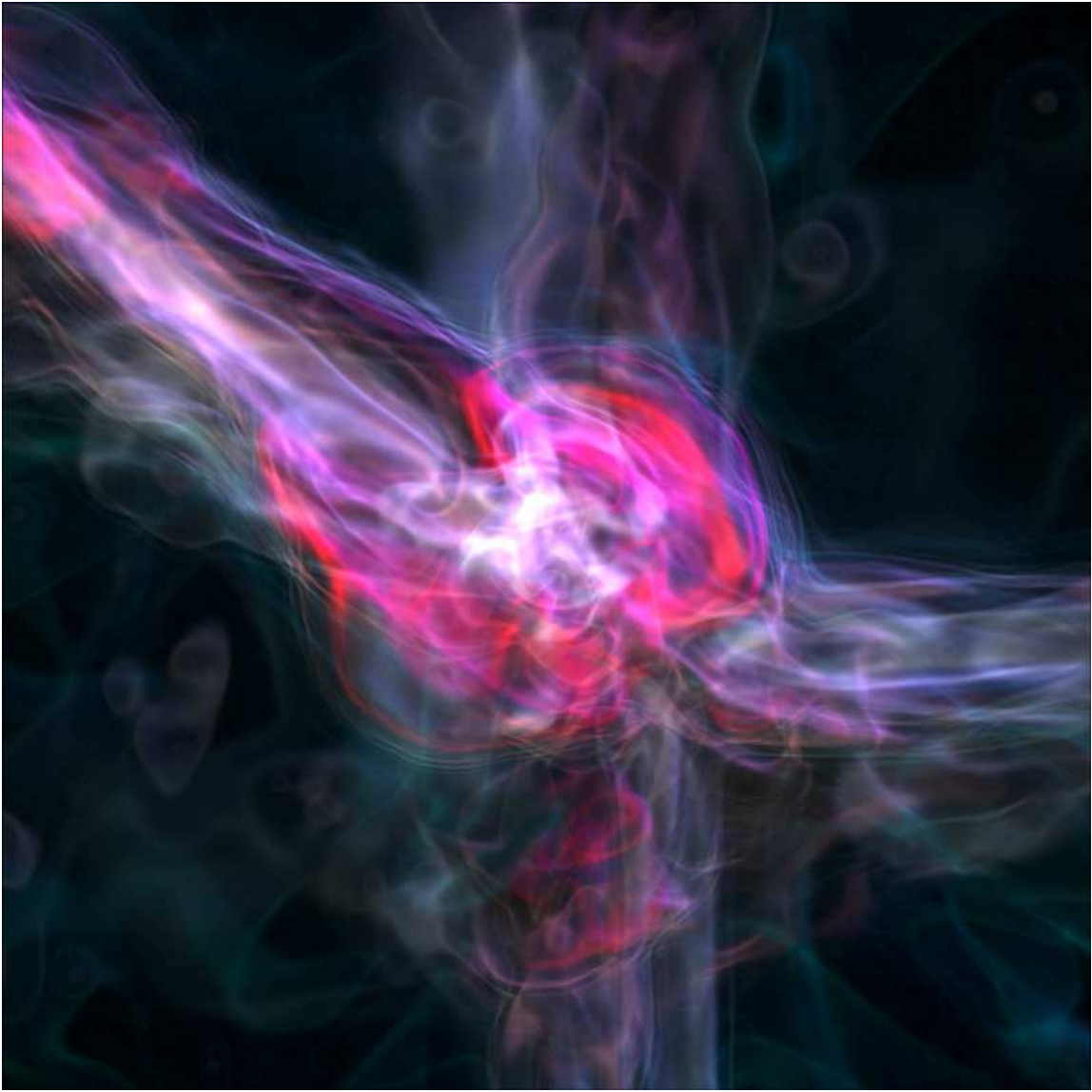}
\caption{Isocontours of temperature for a numerical cluster.  The field of view is 13.85 $h^{-1}$ Mpc.  The colors correspond to the following temperatures: cyan = $10^5$ K, magenta = $3 \times 10^5$ K, blue = $10^6$ K, orange = $3 \times 10^6$ K, red = $1.6 \times 10^7$ K, and white = $5 \times 10^7$ K.  This new visualization was created using a ray casting module recently developed within \textit{yt}.  This image was made by first defining a red-green-blue-alpha transfer function based on any grid quantity.  These emissivities are then integrated along a given ray from back to front.  Note the complex interplay in temperature between the filaments, the cluster periphery, and cluster core.}
\label{Timage}
\end{center}
\end{figure}

Second, this accretion also leads to bulk and turbulent gas motions in the ICM. The bulk gas velocities are often hundreds to even thousands of km/sec which may be detectable with upcoming high resolution X-ray calorimeter spectrometers (\eg Astro-H XCS).  One symptom of these bulk velocities and turbulent motions is the complex asymmetric temperature structures which are visible in simulated clusters as shown in Figures \ref{ImageGrid} and \ref{Timage}.  The particular visualization in Figure \ref{Timage} was chosen to display thin isocontours of temperature, highlighting the various phases of the intracluster gas.  This temperature image illustrates the complex interplay between gas of different temperatures (and velocities) accreting from filaments at multiple angles and thermalizing via a web of shocks throughout the cluster. 

\begin{figure}[t]
\begin{center}
\vspace{-1.0cm}
\includegraphics[width=6.0in]{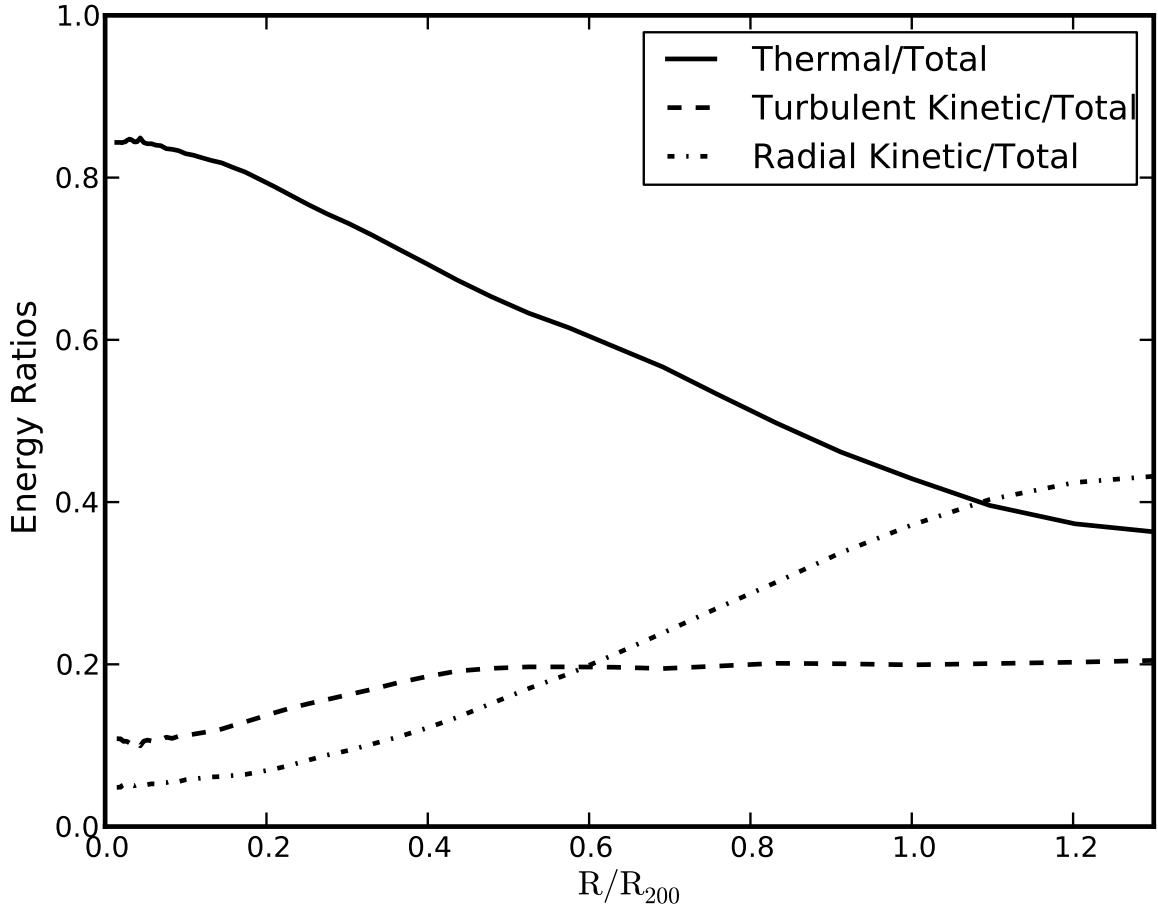}
\caption{Energy density ratio profiles for the sample of numerical clusters.  The thermal, radial kinetic, and turbulent kinetic energy densities are normalized to the total energy density within each shell.}

\label{energies}
\end{center}
\end{figure}

Third, the kinetic gas motions are a significant fraction of the total energy density of the ICM, especially in the outer reaches of clusters.  In Figure \ref{energies}, we plot the radial profiles of the ratios of the thermal ($nkT$), radial kinetic ($\rho v_r^2$), and turbulent kinetic energy (= difference between total and radial kinetic energy) densities to that of the total energy density of the gas in our numerical clusters. To calculate the kinetic energy density of the gas, we first find the
center of mass velocity of the halo.  We then compute the difference
between a cell's velocity and this center of mass velocity.  This then gives
the kinetic motion relative to the halo.  Both the kinetic energy and
thermal energy are calculated using an X-ray emission weighting.  Near the cluster centers ($\lesssim 0.2 r_{200}$), the radial kinetic and turbulent kinetic energies contribute only $\approx 15$\% to the total energy of the ICM, consistent with mild biasing of the hydrostatic mass estimates in the central regions of clusters.  However, at $r_{200}$, the radial kinetic energy equals the thermal energy.  Furthermore, the turbulent energy density at the cluster edge is now $\approx 50$\% of the thermal energy and, thus, provides significant pressure support.  So, it is not surprising that hydrostatic equilibrium increasingly fails as a valid assumption as we get closer to the cluster periphery.  Turbulent gas motion is an important ingredient in the pressure support of galaxy clusters.  For Equation (2) to result in an accurate estimate of cluster mass, it must contain a dynamical gas pressure term ($\sim$$\rho v^2$) \citep[see also][]{2009ApJ...705.1129L}.

\clearpage
\clearpage

\clearpage

\section{Summary and Conclusions}\label{sec:summary}

In comparing new \textit{Suzaku} X-ray observations out to $\approx r_{200}$ for two rich galaxy clusters, PKS 0745-191 and A1795, with numerical clusters extracted from a large cosmological volume, we find that the observed clusters follow universal temperature, density, and entropy profiles.  This excellent agreement between observations and simulations suggests that the simple adiabatic gas physics of the simulations appears to adequately define the thermodynamics outside the cores of these clusters.  With the present \textit{Suzaku} X-ray data, cooling, non-gravitational heating (via \eg AGNs), magnetic fields, or cosmic ray pressure effects are not required.

These profiles, however, are not consistent with simple hydrostatic equilibrium  in the outer cluster regions, suggesting that the ICM pressure support does not arise solely from the thermal energy density.  Using cosmological numerical simulations, we show that nonspherical accretion from large-scale filaments and the resulting gas motions (radial and turbulent kinetic) have a comparable energy density to the thermal energy density near $r_{200}$.  Assuming that thermal gas motions provides all the pressure support results in an increasingly inaccurate estimate of the total cluster mass with radial distance from cluster centers.

\acknowledgements{We thank Marc Bautz, Eric Miller, and Matt George for sharing their electronic files of the X-ray profiles for A1795 and PKS 0745-191, respectively.  We also thank Eric Hallman, Britton Smith, Marc Bautz, Richard Mushotzsky, Eric Miller, and Megan Donahue for helpful conversations.  Simulations were performed in part at the Los Alamos National Laboratory under the auspices of the Institutional Computing Program.  J.O.B. acknowledges support from NASA ADP grant NNZ07AH53G and NSF grant AST-0807215.  S.W.S. has also been supported by a DOE Computational Science Graduate ~Fellowship under grant number DE-FG02-97ER25308.  B.W.O. has been supported in part by a grant from the NASA ATFP program (NNX09AD80G). Computations described in this work were performed using the \textit{Enzo} code developed by the Laboratory for Computational Astrophysics at the University of California in San Diego (http://lca.ucsd.edu)}.

\bibliography{Tedge}
\end{document}